\documentclass[manuscript, screen]{acmart} 

\usepackage [english]{babel}
\usepackage [autostyle, english = american]{csquotes}
\MakeOuterQuote{"}

\usepackage[inline]{enumitem} 

\usepackage{tagging}

\usepackage{hhline}
\usepackage{soul}

\colorlet{soulblue}{cyan!40}
\colorlet{soulred}{red!30}

\AtBeginDocument{%
  \providecommand\BibTeX{{%
    \normalfont B\kern-0.5em{\scshape i\kern-0.25em b}\kern-0.8em\TeX}}}

 \setcopyright{acmcopyright}
\copyrightyear{2021}
\acmYear{2021}

\acmConference[SimuRec'21]{SimuRec '21: Workshop on Simulation Methods for Recommender Systems at ACM RecSys '21}{October 2nd, 2021}{Online, Worldwide}
\acmBooktitle{SimuRec'21: Workshop on Simulation Methods for Recommender Systems at ACM RecSys'21, October 2nd, 2021, Online, Worldwide}

\makeatletter
\renewcommand\@formatdoi[1]{\ignorespaces}
\makeatother

\begin{document}

\title[Towards Creating a Standardized Collection of Simple and Targeted Experiments for RS]{Towards Creating a Standardized Collection of Simple and Targeted Experiments to Analyze Core Aspects of the Recommender Systems Problem}

\author{Andrea Barraza-Urbina}
\affiliation{%
  \institution{Insight Centre for Data Analytics, Data Science Institute, NUI Galway}
  \city{Galway}
  \country{Ireland}
 }
\email{andrea.barraza@insight-centre.org}
\renewcommand{\shortauthors}{Barraza-Urbina}

\begin{abstract}
Imagine you are a teacher attempting to assess a student's level in a particular subject. 
If you design a test with only hard questions, and the student fails, this mostly proves that the student does not understand the more advanced material. 
A more insightful exam would include different types of questions varying in difficulty to truly understand the student's weaknesses and strengths from different perspectives. 
In the field of Recommender Systems (RS), more often than not, we design evaluations to measure an algorithm's ability to optimize goals in complex scenarios, representative of the real-world challenges the system would most probably face. 
Nevertheless, this paper posits that testing an algorithm's ability to address both simple and complex tasks/problems would offer a more detailed view of performance to help identify, at a more granular level, the weaknesses and strengths of solutions when facing different scenarios/domains. 
We believe the RS community would greatly benefit from creating a collection of standardized, simple, and targeted experiments, which, much like a suite of "unit tests", would individually assess an algorithm's ability to tackle core challenges that make up complex RS tasks.
What's more, these experiments go beyond traditional pass/fail "unit tests".
Running an algorithm against the collection of experiments allows a researcher to empirically analyze in which type of settings an algorithm performs best and \textit{to what degree} under different metrics.
Not only do we defend this position, in this paper, we also offer a proposal of how these simple and targeted experiments could be defined and shared and suggest potential next steps to make this project a reality.



\end{abstract}



\maketitle

\section{Introduction}

Recommender Systems (RS) offer a mix of services that aim to imitate, support and/or augment the social process of creating and sharing recommendations~\cite{Terveen2001}. 
To fully appreciate the range of possible applications, it is convenient to think about the everyday situations where people request and offer recommendations.
For instance, when faced with a decision, we usually turn to people we trust, who know us best, to seek personalized advice.
Also, we often feel the need to proactively share our opinions about a positive or negative experience to inform others.
RS enable these processes at a large scale: they aim to learn from the mass of available user opinions and information sources to offer customized suggestions to help users discover and make decisions about items.
%
By reflecting on the numerous ways to exploit a collection of user opinions, it is no surprise that
a broad spectrum of RS services exists for various use cases and application domains, such as movies, news, fashion, restaurants, education, health and lifestyle, among many others.

Evaluating a RS solution/algorithm can be rather challenging.
At a high level, an evaluation consists of one or more experiments. In each, we assess the RS's ability to address some class of problems/tasks quantified by a set of performance measures%
%
\footnote{According to Mitchell~\cite{mitchell_mlbook}: \textit{"In general, to have a well-defined learning problem, we must identity these three features: the class of tasks, the measure of performance to be improved, and the source of experience."}}.
In many ways, an evaluation is similar to a school exam, aiming to determine a student's level in a particular subject.
For a student's performance on the exam to represent their actual knowledge, an exam generally includes a comprehensive set of different types of questions. 
Not only do questions need to cover the relevant topics, but they also need to vary in difficulty to truly understand the student's weaknesses and strengths from different perspectives.
In RS, more often than not, we test algorithms on their ability to optimize goals in complex scenarios, representative of the real-world challenges the system would most probably face.
In short, we tend to create evaluations/exams for RS with only "hard" tasks/questions.
For instance, a "hard" task might be for the RS to optimize several metrics~(such as accuracy, diversity, coverage~\cite{kaminskas2016diversity}), in a complex simulation, where the system would have to offer item suggestions from an extensive and dynamic catalogue to a frequently changing pool of users on different devices.
In contrast, a "simple" task might be for the RS to offer suggestions to a small and static set of known users from a small and static item catalogue to increase rating prediction accuracy.

This paper posits that testing an algorithm's ability to address both simple and complex tasks/problems would offer a more detailed view of performance to help identify, at a more granular level, the weaknesses and strengths of solutions when facing different scenarios/domains.
The ideas presented in this paper are heavily inspired by the work of Osband~et~al.~in~\cite{Osband2020Behaviour}, where the authors introduce a \textit{"Behaviour Suite for Reinforcement Learning"} (more in the next section). In fact, our main argument is that the RS community would benefit from the creation of our own \textit{"Behaviour Suite"}: a collection of standardized, \textit{simple}, and \textit{targeted} experiments, which, much like a suite of "unit tests", would individually assess an algorithm's ability to tackle core challenges that make up complex RS tasks.
What's more, these experiments go beyond traditional pass/fail "unit tests".
Running an algorithm against the collection of experiments allows a researcher to empirically analyze in which type of settings an algorithm performs best and \textit{to what degree} under different metrics.
The following sections will provide more details on the proposed approach and highlight arguments in favour and against it.
Finally, we lay out potential next steps to make the presented ideas a reality.

\section{Proposed Approach}

The \textit{"Behaviour Suite for Reinforcement Learning"} (or simply \texttt{bsuite}) 
is an open source library of \textit{"carefully-designed experiments that investigate core capabilities of reinforcement learning~(RL) agents"}~\cite{Osband2020Behaviour}.
\texttt{bsuite} aims to play a role similar to that of the `MNIST' dataset which \textit{"offers a clean, sanitised, test of image recognition as a stepping stone to advanced computer vision"}~\cite{Osband2020Behaviour}.
The authors highlight that a collection of clear experiments, where each embodies a fundamental issue of the Reinforcement Learning~(RL) problem, would be a powerful driver for progress.
Among five key qualities, experiments in \texttt{bsuite} must be \textit{targeted}, in that performance on the task corresponds to a single key issue in RL, and \textit{simple}, in that it \textit{"it strips away confounding/confusing factors in research"}. 

So far, we have compared designing an evaluation for a RS approach to a teacher creating a school exam.
In this case, \texttt{bsuite}-type experiments would be analogous to what we have called "simple" questions in the exam.
Moving forward, we develop the general idea of creating simple and targeted experiments for RS from a RL point of view.
For this, we first briefly introduce RL and RL-based RS.
Though we focus on a RL perspective, the underlying messages should also be relevant to experiments based on other paradigms used to represent the RS problem, such as Supervised Learning.

\begin{figure}[t]
\centering
\includegraphics[scale=0.53]{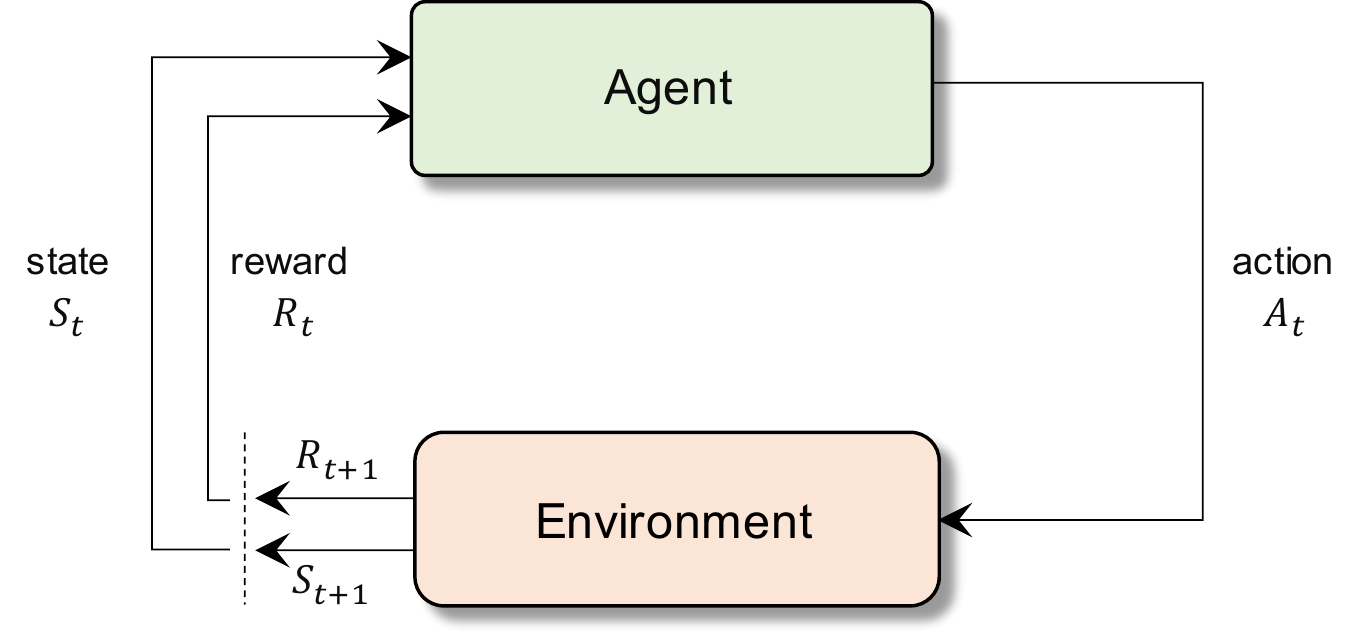}
\caption{The Reinforcement Learning Framework (Source:~\cite{rl_book}).}
\label{fig:rl_framework}
\end{figure}

Reinforcement Learning~(RL) is a branch of Machine Learning inspired by how humans and animals learn from trial-and-error experiences. 
In RL, a goal-oriented \textit{agent} must learn to perform a task despite uncertainty about its \textit{environment}. 
The agent does not receive examples or instructions of the correct actions to perform in different situations.
Instead, in response to actions,
the agent obtains feedback from the environment in the form of positive or negative \textit{reward signals}.
Positive rewards are meant to reinforce actions/behaviour and negative rewards are meant to discourage actions.
%
%
%
The interaction between the agent and its environment is formalized by the \textit{RL Framework} in~Figure~\ref{fig:rl_framework}, an \textit{"abstract and flexible"} framework that can be used to represent different problems where a decision-making agent needs to learn from interaction~\citep{rl_book}.
For us, the relevant aspects of the framework are:%
\begin{itemize}[topsep=1pt]
    \item The \textit{agent} is the learner or decision-maker. 
    \item The \textit{environment} characterizes the agent's application domain and the task/goals the agent is trying to achieve.
    \item The agent interacts with the environment by performing actions in a continuous sequence of discrete time steps.
    \item The state represents all the information about the environment available to the agent at a specific time step.
    \item A reward is a numeric score that is used to motivate the agent to learn how to perform actions in different states/situations.
    \item From the agent's perspective, the environment has a simple interface reduced to the information found in states, actions and rewards.
\end{itemize}

Conventionally, the RS problem has been formalized as a Supervised Learning task (Batch Learning). However, recent works have proposed that RL can be a more appropriate paradigm to frame the “modern” and Interactive RS problem~(more details in~\cite{barraza2018bears}). In an RL-based RS: The RS algorithm is viewed as a sequential decision-making agent focused on learning over time how to perform actions (e.g.,~offer item suggestions), in different states (e.g.,~to different users), using interactive feedback in the form of reward signals (e.g.,~user ratings). The RS problem/task is characterized by the environment.
%
Although, at first glance, the environment could be confused with a simulation,
in~\cite{barraza2019towards} we provide more details on why datasets and simulations
require additional assumptions to specify a RS task fully.
To simplify, the reader can interpret the environment as a special type of simulation that requires no additional assumptions to generate a state and a reward in response to an agent's action.

Now that we have provided the necessary background, we can continue defining what a simple and targeted experiment would be.
In \texttt{bsuite}, an \textbf{\textit{experiment}} has three components:

\begin{enumerate}[topsep=1pt]
    \item \textbf{Environments}: A fixed set of environments that correspond to a targeted/specific type of problems/tasks. As a reminder, a \textit{targeted} task corresponds to a single key issue in RL.
    \item \textbf{Interaction}: A fixed strategy defining how the agent-environment interact. For instance, we can define that the agent and environment interact for a $100$ time steps. Another relevant parameter is the number of \textit{episodes}, where an episode is
    a sequence of interactions starting from an initial time step and ending on a final time step. Due to the stochastic nature of both the agent and environment, it is helpful to measure the agent's expected performance when interacting with the environment over several episodes.
    Interesting examples of interaction can be found in~\cite{whiteson2010report}\footnote{The authors do not explicitly use the term interaction, but use the terms \textit{experiment} and \textit{evaluation}.}~(e.g.,~interactions with training and testing phases).
    For a concrete example in RS, see~\cite{barraza2018bears}\footnote{The author uses the term \textit{experiment} instead of interaction.}.
    
    \item \textbf{Analysis}: \textit{"a fixed procedure that maps agent \textit{behaviour} to results and plots"}~\citep{Osband2020Behaviour}. Having an explicit analysis component allows to have a shared understanding of what it means to solve a task and share benchmarks.
    
\end{enumerate}

One example experiment included in \texttt{bsuite} is called \textit{Mountain Car}, a classic RL task where the agent needs to learn to drive an underpowered car up a hill.
To solve the task, the agent must drive back and forth to gain the momentum needed to reach the top of the hill~(see~~\cite[Fig.~2]{whiteson2010report}).
Among challenges, the task represents the issue of delayed reward or credit assignment~\cite{whiteson2010report, Osband2020Behaviour}.
In Mountain Car, the environment provides the agent with constant negative rewards at each time step and only generates a positive reward if the car reaches the top of the hill. This reward scheme makes it difficult for the agent to learn which sequence of actions actually lead to the goal~\cite{whiteson2010report}.

With \texttt{bsuite}, a researcher is in a better position to analyze the weaknesses and strengths of a RL solution in relation to key challenges of a RL problem.
For example,~\cite[Fig.~5]{Osband2020Behaviour} presents a `radar plot' summarizing the agent's performance on several experiments. Each dimension in the plot is a single score quantifying the agent's ability to address a core challenge.
\looseness=-1

In RS, we believe the community has mainly focused on designing evaluations with complex experiments, where in each, the RS is faced with several core challenges simultaneously.
For this, the community has created environments/simulations~\cite{recogym, recsim, mahmood2007learning} and datasets~\cite{harper2015movielens} to share the complexities of specific RS problems/use cases.
Moreover, every year, different competitions are hosted to encourage the community to solve real-world tasks or challenges~(such as the ACM RecSys Challenge~\cite{anelli2020recsys}).
All these efforts are essential to drive progress in the field.
However, we believe an effort is also required to create simple and targeted experiments representative of core challenges in the field, such as user/item cold start, high churn, delayed feedback, privacy, security, trust, fairness, multiple stakeholders, and others.
These simple experiments would serve as complementary tools to existing evaluations. Although they would play a different role than complex experiments, they can have several advantages, as discussed in the following section.


\section{Discussion and Next Steps}

Creating a shared collection of simple and targeted experiments can have several advantages for the RS community. We highlight the following:

\begin{itemize}[topsep=1pt]

    \item \textit{A clear and standardized methodology to evaluate a RS algorithm from different perspectives/dimensions:} The approach encourages a standardized evaluation methodology, much more specific than just sharing an environment/simulation/dataset. This is important to compare and reproduce evaluations across multiple works.

    \item \textit{Deeper understanding of the RS problem:} A \texttt{bsuite} for RS would help the community to have a shared, clear and comprehensive understanding of the core challenges that, as a community, we should be tackling. It would also help identify which essential problems might not be thoroughly defined.
    
    \item \textit{Deeper understanding of the state of the art:} By sharing experiment results in a consistent and comparable way, the community can better understand the state of the art in solving the key challenges of the field. It would also be easier to understand which challenges are better addressed compared to others.
    
    \item \textit{Deeper understanding of algorithm properties and the type of tasks they can solve:} 
    All RS solutions are not well suited for all RS tasks~\cite{burke2011matching, adomavicius2012impact}. The challenge of matching RS techniques to RS problems was explored by Burke and Ramezani~in~\cite{burke2011matching}. Their approach was to analyze the underlying properties of different application domains~(such as,~news, e-commerce, music, and others) to provide a matching between domains and RS technologies~(e.g.,~a Collaborative Filtering approach would be better suited for a movie RS task). With a \texttt{bsuite} for RS, it would be easier to empirically analyze in which type of settings an algorithm performs best by running the algorithm against the collection of experiments. This process would not have to depend on the type of domain, as it might be the case that different RS tasks in the same domain might have different underlying core challenges.

    \item \textit{"Unit Testing" for faster and more informed development of RS solutions:} \texttt{bsuite}-type experiments can help diagnose faster problems during the development of a RS solution. The fact that the researcher would have these experiments readily available, would mean faster development as well.

    \item \textit{Shared Benchmarks:} Having a collection of standardized experiments means we can have shared benchmarks. Among advantages, shared benchmarks can help to identify the best algorithms for a specific core challenge.
    

\end{itemize}

Arguments against the approach could include:

\begin{itemize}[topsep=1pt]
    \item \textit{Would using a comprehensive set of metrics not be enough to achieve the same level of analysis as using a collection of simple experiments?} Multiple metrics exist 
    for different RS goals~\cite{kaminskas2016diversity}. Using several metrics is part of the \textit{Analysis} component of an experiment. We believe that using a comprehensive set of metrics as part of a collection of diverse experiments would provide more information regarding an agent's performance.
    
    \item \textit{Designing a targeted experiment for a single core challenge may prove to be too hard:} Even the most basic RS task might require for the solution to address several challenges. Even in~\cite{Osband2020Behaviour}, presented experiment examples were associated with two or three key RL issues. The definition of `targeted' might need to be revisited and adapted to what it would mean for the RS community.  
    
\end{itemize}

Making the proposed approach a reality would entail several challenges. We highlight the following:
\begin{itemize}[topsep=1pt]
    \item  \textit{Community effort to create experiments:} Designing experiments to represent an algorithm's ability to address a core challenge can prove to be an intricate task and would require the engagement of several experts from the many "sub-fields" of RS. Finding incentives to motivate the community would prove challenging if the project does not gain support from community leaders. A possible solution would be to form a small team to create an initial version of the library, such as \texttt{bsuite}. The intuition would be that if researchers find value in the smaller version of the library they would be encouraged to participate in future versions of the project.

    \item \textit{Defining specific criteria experiments have to meet to be added to the library:} To address this challenge, \texttt{bsuite} expects to form a committee that can review and select experiments that have been proposed by the community. The committee is expected to meet annually during the NeurIPS conference.

    \item \textit{Practical/technical aspects of developing and sharing experiments:} \texttt{bsuite} is an open source project developed in Python and mostly maintained by researchers at DeepMind. It needs to be determined who would adopt ownership and general maintenance responsibilities of a future codebase.

\end{itemize}

Given the discussed challenges, we propose the following next steps:
\begin{itemize}[topsep=1pt]
    \item Refine the details of the proposed approach and define specific requirements for an initial version of a \texttt{bsuite} library for RS. The initial version should include a small collection of experiments and example analysis reports~(view appendices in~\cite{Osband2020Behaviour}).
    \item Define incentive mechanisms to motivate the RS community to participate in the project.
    \item Define effective community outreach strategies to raise awareness about the project.
    \item Find sponsors for the project willing to take ownership and develop the first version.
    \item Form a committee that can review and select experiments submitted by the community.
\end{itemize}

\begin{acks}
Special thanks to Brian Tanner for introducing the author to \texttt{bsuite} and for the interesting discussions.
This publication has emanated from research conducted with the financial support of Science Foundation Ireland (SFI) under Grant Number SFI/12/RC/2289\_P2, co-funded by the European Regional Development Fund. For the purpose of Open Access, the author has applied a CC BY public copyright licence to any Author Accepted Manuscript version arising from this submission.
\end{acks}

\bibliographystyle{ACM-Reference-Format}
\bibliography{references}

\end{document}